\documentclass[runningheads]{llncs}

\usepackage{amsmath}
\usepackage{oz}
\usepackage{graphicx}
\usepackage[english]{babel}
\usepackage[T1]{fontenc}
\usepackage{tikz-cd}
\usepackage{caption}
\usepackage{framed}
\usepackage{subcaption}
\usepackage[utf8x]{inputenc}
\usepackage{amssymb, upgreek}

\usepackage{color}

\usepackage{rotating}

\definecolor{keywordcolor}{rgb}{0.7, 0.1, 0.1}   
\definecolor{commentcolor}{rgb}{0.4, 0.4, 0.4}   
\definecolor{symbolcolor}{rgb}{0.0, 0.1, 0.6}    
\definecolor{sortcolor}{rgb}{0.1, 0.5, 0.1}      
\definecolor{errorcolor}{rgb}{1, 0, 0}           
\definecolor{stringcolor}{rgb}{0.5, 0.3, 0.2}    

\usepackage{listings}
\usepackage{chngcntr}

\lstdefinestyle{lean}{
	float=t,
	numbers=left
}

\usepackage{caption}
\usepackage{subcaption}
\usepackage{proof}
\usepackage{algorithm2e}
\usepackage{color}

\newcommand{\term}[1]	{\emph{\textcolor{teal}{#1}}}

\newcommand{\fig}[1]		{Fig.~\ref{fig:#1}}

\newcommand{\BD}{\begin{definition}}
\newcommand{\ED}{\end{definition}}

\newcommand{\TT}{\text{True}}
\newcommand{\FF}{\text{False}}

\newcommand{\tuple}[1]{\langle #1 \rangle}

\newcommand{\name}[1]{$\mathtt{#1}$}
\newcommand{\code}[1]{\texttt{#1}}
\DeclareMathOperator{\Prop}{Prop}
\DeclareMathOperator{\Conf}{Conf}

\newcommand{\SPL}       {\mathcal{L}}
\newcommand{\featset}{F}
\newcommand{\featmodel}{\Phi}
\newcommand{\domainmodel}{D}
\newcommand{\pcmap}{\phi}
\newcommand{\config}{\rho}
\newcommand{\PC}[1]{\texttt{#1}}
\newcommand{\indexProd}[2]{#1|_#2}
\begin{document}
\counterwithout{lstlisting}{chapter}

\title{Towards Certified Analysis of Software Product Line Safety Cases}
\author{Ramy Shahin\inst{1} \and
	Sahar Kokaly\inst{2} \and
	Marsha Chechik\inst{1}}
\institute{University of Toronto, Toronto, Canada\\
	\email{\{rshahin,chechik\}@cs.toronto.edu} \and
	General Motors, Canada\\
	\email{sahar.kokaly@gm.com}}
\maketitle

\vspace{-0.2in}
\begin{abstract}
	Safety-critical software systems are in many cases designed and implemented as families of products, usually referred to as Software Product Lines (SPLs). Products within an SPL vary from each other in terms of which features they include. Applying existing analysis techniques to SPLs and their safety cases is usually challenging because of the potentially exponential number of products with respect to the number of supported features.
	In this paper, we present a methodology and infrastructure for certified \emph{lifting} of existing single-product safety analyses to product lines. To ensure certified safety of our infrastructure, we implement it in an interactive theorem prover, including formal definitions, lemmas, correctness criteria theorems, and proofs.
	
	We apply this infrastructure to formalize and lift a Change Impact Assessment (CIA) algorithm. We present a formal definition of the lifted algorithm, outline its correctness proof (with the full machine-checked proof available online), and discuss its implementation within a model management framework. 
\end{abstract}
\vspace{-0.2in}

\vspace{-0.2in}
\section{Introduction}
\label{sec:intro}
\vspace{-0.1in}
The development of safety-critical systems usually involves a rigorous safety engineering process. A primary artifact resulting from that is a \term{safety case}, identifying potential safety hazards, their mitigation goals, and pieces of evidence required to show that goals have been achieved.
Safety cases, together with other system artifacts, are usually inspected and analyzed by tools as a part of the safety engineering process. In safety-critical domains, correctness of those tools is essential to the integrity of the whole process. \term{Correctness certification} of tools w.r.t. their specifications becomes of extremely high value in this context.

In many cases, families of safety-critical software products are developed together in the form of  \term{Software Product Lines (SPLs)}. Different product variants of an SPL have different \term{features}, i.e., externally visible attributes such
as a piece of functionality, support for a particular peripheral device, or a performance optimization. Each feature can be either present or absent in each of the product variants of an SPL.
Given this combinatorial nature of feature composition, 
analyzing the safety of each product instance individually in a \term{brute-force} fashion is usually intractable. 

Several source-code and model-based analysis tools have been \term{lifted} to product lines~\cite{Thum:2014,Kastner:2012,Gazzillo:2012,Salay:2014,Bodden:2013,Classen:2013,Shahin:2019,Shahin:2020} in the sense that they can be applied efficiently to the whole product line at once, leveraging the commonalities between individual products, and thus generating aggregated results for the complete set of products. Those results have to be correct with respect to applying the analysis to each product individually. 
However, to the best of our knowledge, lifting of safety analyses has not been attempted before. 

In this paper, we present a systematic methodology to correct-by-construction lifting of safety case analysis algorithms to software product lines. This includes infrastructure building blocks for implementing lifted algorithms, and proving their correctness with respect to their single-product counterparts. We use the Lean interactive theorem prover~\cite{deMoura:2015} to formalize the correctness criteria of lifting, implement our lifting infrastructure, and prove the correctness of lifted algorithms. A Lean proof is machine-checked, so it can be used as a \term{correctness certificate} of the property being proven.

We demonstrate our approach on a \term{Change Impact Assessment (CIA)} algorithm~\cite{Kokaly:2017} that takes a system model, an assurance case, traceability links in between, and a modification to the system model as inputs, and determines the set of safety case elements that need to be revised or rechecked.

\begin{figure}[t]
	\centering
	\includegraphics[width=\textwidth]{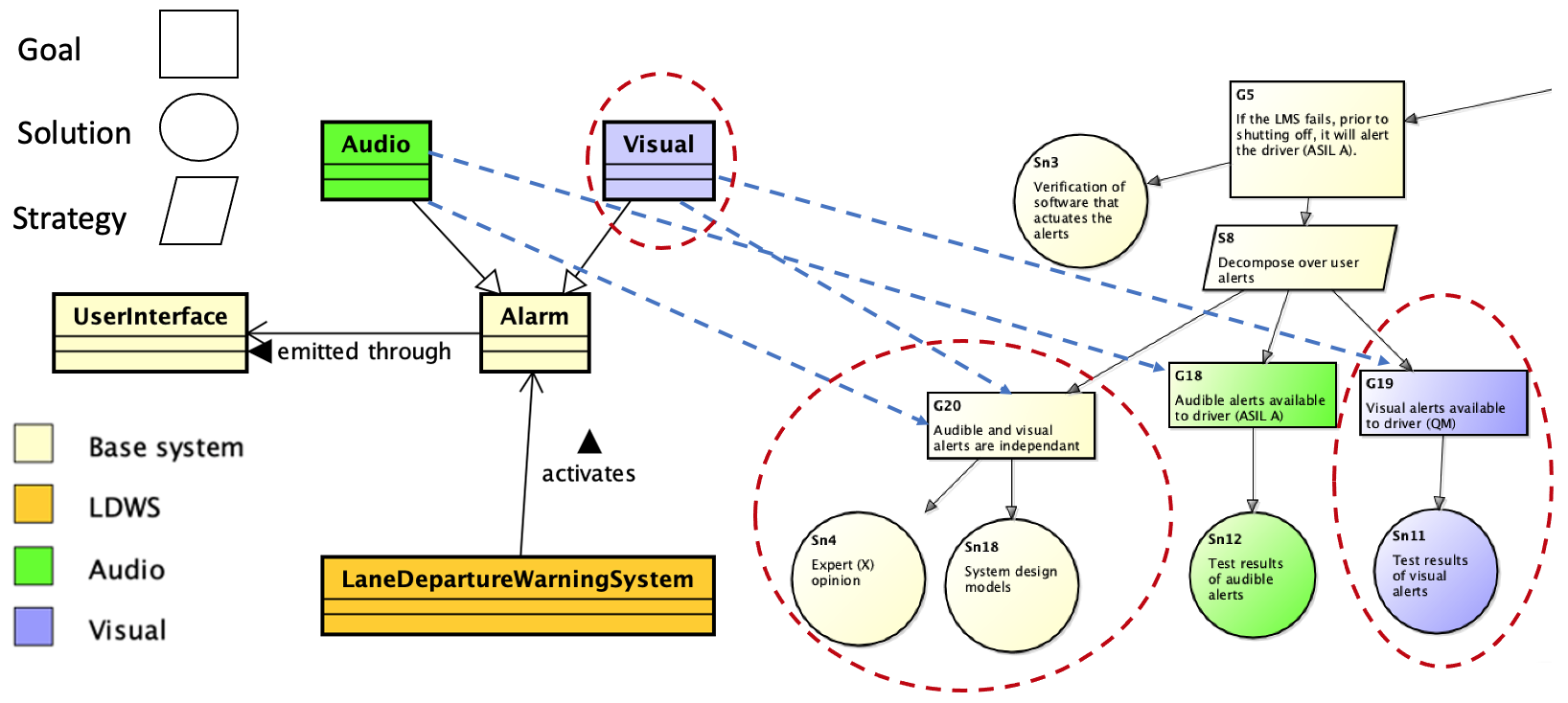}
	\caption{Lifted change impact assessment when the "Visual" class is modified. A dashed ellipse around the Visual class denotes a modification, and dashed ellipses around safety case elements indicate that they need to be rechecked as a result.}
	\label{fig:ex}	
	\vspace{-0.3in}
\end{figure}

\vskip 0.1in
\noindent
{\bf Motivating Example.}
Consider the \term{Lane Management System (LMS)} system outlined in~\cite{Chechik:2020}. 
LMS can be thought of as a product line with several features, including: \term{Lane Departure Warning System (LDWS)}, \term{Audio warning (Audio)}, and \term{Visual warning (Visual)}. For simplicity of presentation, we assume that all feature combinations are allowed.
Fig.~\ref{fig:ex} shows a snippet of the class diagram of the LMS product line, and the corresponding snippet of its GSN~\cite{Kelly:2004} assurance case, with traceability links in between the two. 

We use colored annotations to map class diagram and GSN elements to features. For example, elements colored in green belong to the \name{Audio} feature, and those in orange belong to the \name{LDWS} feature. Base system elements (existing in all products) are in yellow. 
In general each element can be annotated by a propositional formula over features (usually referred to as a \term{presence condition}).

Consider a modification to the  \name{Visual} class. The problem CIA algorithms try to solve is figuring out how that modification of a system element would impact the safety case. 
We distinguish between two ways in which a change to the system can impact safety case elements~\cite{Kokaly:2017}: 
(1) \name{revise} -- the \emph{content} of the element (e.g., definition of a goal, or description of a solution) may have to be revised because it referred to a system element that has changed and the semantics of the content may have changed, and 
(2) \name{recheck} -- the \emph{state} of the element (e.g., whether a goal is satisfied, or a solution is available) must be rechecked because it may have changed.

In a product line setting, in addition to figuring out which elements are impacted, we also need to identify the product variants in which they are. In Fig.~\ref{fig:ex}, goals \name{G19} and \name{G20} are directly impacted by modifications to class \name{Visual} because of the direct traceability links. 
Both classes need to be rechecked as a result, but only in products where the \name{Visual} feature is included.
In the same set of products, pieces of evidence linked to those goals (\name{Sn4}, \name{Sn11}, \name{Sn18}) need to be rechecked as well.
Note that although \name{G20}, \name{Sn4}, and \name{Sn18} belong to all product variants, we do not need to recheck them in product variants not including the \name{Visual} feature.

A CIA tool lifted to product lines has to preserve the exact semantics of its single-product counterpart. In other words, using the lifted tool should output exactly the union of outputs of the single-product tool applied to each product variant. A software bug in the lifted tool might result in false positives (elements marked as impacted while they should not). Even worse, a bug might result in overlooking an impacted element, potentially resulting in safety incidents.

\vskip 0.05in
\noindent
{\bf Contributions.}
In this paper we
(1) outline a methodology for lifting safety analyses to safety cases of software product lines, and present a generic infrastructure for certified lifting (data structures and correctness criteria) using the Lean interactive theorem prover; and
(2) demonstrate our methodology on a CIA algorithm lifted to software product lines, i.e., supporting the input of feature-specific modifications, and outputting feature-specific annotations of safety case elements. In addition,
(3) we formalize the single-configuration CIA algorithm from~\cite{Kokaly:2017} using Lean;
(4) we outline a sketch of the correctness proof of the lifted CIA algorithm with respect to the single-configuration one (full Lean proof available online); and
(5) we discuss extending MMINT-A~\cite{Fung:2018} model management framework with lifted safety algorithms, including lifted CIA.

\vskip 0.05in
\noindent
{\bf Organization.}
The rest of this paper is organized as follows:
In Sec.~\ref{sec:background}, we provide background on safety cases and SPLs. 
We outline the correctness criteria, methodology, and infrastructure needed to formally lift safety case algorithms in Sec.~\ref{sec:methodology}.
In Sec.~\ref{sec:cia} we formalize the original single-configuration CIA algorithm, its lifted counterpart, and outline the lifting correctness proof.
Sec.~\ref{sec:impl} explains how lifted algorithms can be integrated into existing model management tools.  Sec.~\ref{sec:related} compares our approach to related work, and Sec.~\ref{sec:conclusion} concludes.

\newcommand{\pf}{\pfun}
\newcommand{\sys}{Sys}		
\newcommand{\sinit}{S}
\newcommand{\snext}{S'}
\newcommand{\ac}{GSN}		
\newcommand{\tr}{TraceRel}	
\newcommand{\Ea}{C_a}		
\newcommand{\Em}{C_m}		
\newcommand{\Ed}{C_d}		
\newcommand{\annot}{K}		
\newcommand{\revi}{revise}
\newcommand{\rec}{recheck}
\newcommand{\reu}{reuse}

\section{Background}
\label{sec:background}
\vspace{-0.1in}

\subsection{Safety Cases, GSN, and Change Impact Assessment}
\vspace{-0.1in}
A safety case is a structured argument, decomposing safety goals into sub-goals, and linking pieces of safety evidence to the goals. Safety goals are usually identified using hazard assessment techniques. Each of the hazards needs to be mitigated by fulfilling one or more safety goal(s).

Goal Structured Notation (GSN)~\cite{Kelly:2004} is a graphical notation for defining safety cases. The safety case portion of Fig.~\ref{fig:ex} is an example of a GSN safety case model. A GSN model has elements of four different types. 
A \term{goal} is either satisfied or not based on the states of its sub-goals, connected solution nodes, and the semantics of decomposition strategy nodes involved. A \term{solution} is a piece of evidence that needs to be validated for its connected goal(s) to be satisfied. A \term{strategy} is a decomposition of a goal into sub-goals. A \term{context} connected to a goal node adds contextual assumptions that are assumed to hold when evaluating whether a goal is satisfied or not.

GSN-IA~\cite{Kokaly:2017} is an algorithm for reflecting changes made in system models onto the relevant GSN safety cases. 
The inputs to GSN-IA are the initial system model $S$
and a safety case $A$ connected by a traceability mapping $R$, the
changed system $S'$ and the delta $D$ recording the changes between
$S$ and $S'$. Specifically, $D$ is the triple $\tuple{C0a, C0d, C0m}$ where $C0a$, $C0d$, and $C0m$ are the 
set of elements added, deleted, and modified respectively.
The output of GSN-IA is the annotated model $K$ of the safety case $A$, indicating which elements are marked for \name{revise},
\name{recheck}, or \name{reuse}.

GSN-IA is parameterized by three slicers~\cite{Salay:2016}: a system model slicer $\mathtt{Slice}_{Sys}$, and two safety case slicers $\mathtt{Slice}_{GSN_V}$ and $\mathtt{Slice}_{GSN_R}$. $\mathtt{Slice}_{Sys}$ is used to
determine how the impact of modifications propagates within the system model. Similarly, the safety case slicers trace through dependencies within the safety case, with $\mathtt{Slice}_{GSN_V}$ only tracing direct dependencies, while $\mathtt{Slice}_{GSN_R}$ recursively generates the transitive closure of dependencies.

\vspace{-0.15in}
\subsection{Software Product Lines}
\vspace{-0.1in}
We introduce Software Product Line (SPL) concepts following definitions from~\cite{Salay:2014}.
An SPL $\SPL$ is a tuple $(\featset, \featmodel, \domainmodel, \pcmap)$ where:
(1) $\featset$ is the set of features s.t.  an individual product can be derived from $\SPL$ via a \term{feature configuration} $\config \subseteq \featset$.
(2) $\featmodel \in \Prop(\featset)$ is a propositional formula over $\featset$ defining the valid set of feature configurations. $\featmodel$ is called a \term{Feature Model (FM)}. The set of valid configurations defined by $\featmodel$ is called $\Conf(\SPL)$.
(3) $\domainmodel$ is a set of program elements, called the \term{domain model}. The whole set of program elements is usually referred to as the \term{150\% representation}. 
(4) $\pcmap:\domainmodel \to \Prop(\featset)$ is a total function mapping each program element to a proposition (\term{feature expression}) defined over the set of features $F$. $\pcmap(e)$ is called the \term{Presence Condition (PC)} of element $e$, i.e. the set of product configurations in which $e$ is present.

Given a product line $\SPL$ and a feature configuration $\config$, we define $\indexProd{\SPL}{\config}$ to be the subset of elements of $\SPL$ that belong to at least one of the features in $\config$. We loosely use the same indexing operator when referring to subsets of values in a data structure subject to a feature configuration. For example, given a feature configuration \{\name{LDWS, Visual}\}, a product with all the elements, except for the ones annotated in green, is instantiated from the product line (Fig.~\ref{fig:ex}).

\newcommand{\lift}[1] {#1$'$}
\newcommand{\liftt}[1] {#1'}
\newcommand{\find}{\texttt{Find}}
\newcommand{\ff}{\texttt{False}}
\newcommand{\instance}{P}
\newcommand{\result}{R}
\newcommand{\analysis}{f}

\newcommand{\Rn}{R'}
\newcommand{\TR}{\lift{TraceRel}}
\newcommand{\Cza}{C0a}
\newcommand{\Czd}{C0d}
\newcommand{\Czm}{C0m}
\newcommand{\Codm}{C1dm}
\newcommand{\Coam}{C1am}

\newcommand{\tPC}{\code{PC}}
\newcommand{\Var}{\code{Var}}
\newcommand{\lset}{\code{set$'$}}

\vspace{-0.15in}
\section{Methodology and Infrastructure}
\label{sec:methodology}
\vspace{-0.1in}
In this section, we present a set of generic infrastructure building blocks that can be used in designing and certifying the correctness of variability-aware algorithms applied to safety cases.
We then present the correctness criteria of variability-aware algorithms with respect to their single-product counterparts. 
Finally, we put the infrastructure together with the correctness criteria into a correct-by-construction methodology for systematic lifting of safety case algorithms.  

We formalize algorithms, theorems, and proofs using the Lean~\cite{deMoura:2015} interactive theorem prover. 
We had two requirements for the proof assistant to be used in this project: (1) to be based on constructive rather than classical logic, to allow for explicit tracing of which sub-goals (and their proof evidence) contribute to the overall proof; and (2) to allow for sound user-defined proof automation procedures, which can reduce the human effort involved in the proof development process. Lean meets those two requirements. It is based on the Calculus of Inductive Constructions~\cite{Bertot:2010}, so it supports constructive logic by default. It also supports tactic-based meta-programming of theorems and proof objects. 


\begin{lstlisting}[style=lean,basicstyle=\small,firstline=9,lastline=16,caption=Variability-aware building blocks.,label=lst:var]
def PC := Prop
structure Var (α : Type) := (v : α) (pc : PC) 
def set' (α : Type) : Type := α → PC
def index (s : set' α) (pc : PC) : set α := (and pc) ∘ s
def mem (x : Var α) (s : set' α) : Prop := x.pc → (s x.v) 
def subset (s₁ s₂ : set' α) : Prop := ∀ a, mem a s₁ → mem a s₂
def union (s₁ s₂ : set' α) : set' α := λ x, (s₁ x) ∨ (s₂ x)
def image (f : α → β) (s : set' α) : set' β := λ x, (∀ y, f y = x ∧ s y)  
\end{lstlisting}

\vspace{0.1in}
\noindent
{\bf Lifted Data Structures.}
\label{sec:lifted_data}
The types of all input, output, and intermediate data structures of an algorithm need to be lifted, i.e., elements of each of those data structures need to be paired with presence conditions, indicating the set of products this element belongs to. Listing~\ref{lst:var} has definitions of some of the data types used for lifted data structures. \tPC~(line 1) is the type for presence conditions, which is defined as native Lean propositions. \Var~(line 2) is a higher-order lifted type, taking a type $\alpha$ as a parameter, and pairing values of type $\alpha$ with presence conditions. 

The lifted set data type \lset~(line 3) is a higher-order type parameterized by type $\alpha$, and implemented as a function $\alpha \to \tPC$. This implementation happens to be the same as the implementation of Lean sets. However, the semantics of Lean sets assume that a value of type $\alpha$ is either present or absent in a set over $\alpha$. Lifted sets on the other hand map an element of type $\alpha$ to an arbitrary propositional formula which might evaluate to \TT~(i.e., the element exists in the set in all configurations), \FF~(i.e., the element does not exist in any configuration of the set), or a contingent formula indicating the set of configurations in which the element exists in the set.

The primary operation on lifted data types in general is indexing. Given a lifted set \code{s}~and a presence condition \code{pc}, \code{index s pc}~evaluates to a Lean set (not lifted) of elements existing in the configurations satisfied by \code{pc}~in \code{s}. This is exactly how the \code{index} operator is defined on \lset, conjoining \code{pc} with the presence condition of each element in \code{s} (line 4).

Standard set operations also need to be overloaded for lifted sets. Lifted set membership semantically checks if a lifted value \code{(v,pc)} exists in all configurations of a lifted set \code{s}. It is defined (line 5) as a propositional implication between \code{pc} (the set of configurations where the lifted value exists), and the set of configurations where \code{v} exists in \code{s}.

Lifted subset is defined exactly the same as standard subset, using the lifted definition of set membership (line 6). Similarly, lifted set union is implemented as a disjunction of the propositional definitions of its two arguments (line 7). The last lifted set operation is \code{image} (line 8), taking a function $f : \alpha \to \beta$ and a lifted set $s$ of $\alpha$, and applies $f$ to each element $s$, returning a lifted set of $\beta$.

\begin{figure}[t]
	\begin{subfigure}[c]{0.45\textwidth}
		\centering
		\begin{tikzcd}
		\SPL \arrow{r}{\liftt{\analysis}} \arrow{d}{\indexProd{}{\config}} & \liftt{\result} \arrow{d}{\indexProd{}{\config}} \\
		\instance \arrow{r}{\analysis} & \result
		\end{tikzcd}
		\caption{Correctness of a lifted function~\cite{Shahin:2020}.}
		\label{fig:correctness1}
	\end{subfigure}
	\hfill
	\begin{subfigure}[c]{0.55\textwidth}
		\centering
		\begin{tikzcd}
		\SPL \arrow{d}{\indexProd{}{\config}} \arrow{r}{\liftt{\analysis}} & \liftt{\result} \arrow{r}{\liftt{g}} & \liftt{S} \arrow{d}{\indexProd{}{\config}} \\
		\instance \arrow{r}{\analysis} & \result \arrow{r}{g} & S
		\end{tikzcd}
		\caption{Correctness of lifted function composition.}
		\label{fig:correctness2}
	\end{subfigure}
	\vspace{-0.1in}
	\caption{Lifting correctness criteria.}
	\label{fig:correctness}
	\vspace{-0.2in}
\end{figure}

\begin{lstlisting}[style=lean,basicstyle=\small,firstline=42,lastline=46,caption=Lifted function composition theorem.,label=lst:funComp]
variables (f  : set  α → set  β) (g  : set  β  → set  γ)
variables (f' : set' α → set' β) (g' : set' β  → set' γ)
theorem fun_comp_correct :
   (∀ a ρ, (f (a | ρ) = (f' a) | ρ)) → (∀ b ρ, (g (b | ρ) = (g' b) | ρ)) →
   (∀ a ρ, (g ∘ f) (a | ρ) = ((g' ∘ f') a) | ρ) 
\end{lstlisting}

\vspace{0.1in}
\noindent
{\bf Correctness Criteria.}
Given a product line $\SPL$, an analysis algorithm $\analysis$, and a product configuration $\config$, we construct a lifted version of $\analysis$ (referred to as \lift{$\analysis$}), such that instantiating a product $\instance$ from $\SPL$ using configuration $\config$, and then applying $\analysis$ to $\instance$ has the same result as applying \lift{$\analysis$} to $\SPL$ and then instantiating a product-specific result using $\config$. This is summarized by the commuting diagram in Fig.~\ref{fig:correctness1}~\cite{Shahin:2020}.

\vspace{0.1in}
\noindent
{\bf Lifting Methodology.}
We follow a divide-and-conquer methodology to design lifted analyses from their single-product counterparts. If an analysis algorithm is broken-down into smaller functions, and each of those functions is individually lifted, composing the lifted functions together has to preserve the correctness criteria. This is summarized in Fig.~\ref{fig:correctness2}. 

We formulate the correctness criteria of lifted function composition as a theorem (Listing~\ref{lst:funComp}). Assume we have two functions \code{(f:set $\alpha$ $\to$ set $\beta$)} and \code{(g:set $\beta$ $\to$ set $\gamma$)}, and two lifted functions \code{(\lift{f}:\lift{set} $\alpha$ $\to$ \lift{set} $\beta$)} and \code{(\lift{g}:\lift{set} $\beta$ $\to$ \lift{set} $\gamma$)} (Line 2). The theorem states that if \code{\lift{f}} is a correct lifting of \code{f}, and \code{\lift{g}} is a correct lifting of \code{g}, then \code{\lift{g} $\circ$ \lift{f}} is a correct lifting of \code{g $\circ$ f} (Lines 3-5). The theorem is proven by term rewriting. Definitions of all theorems, lemmas, and their full Lean proofs are available online\footnote
{\url{https://github.com/ramyshahin/variability}}.

Correctness of the lifted function composition theorem is the foundation of compositional lifting correctness proofs. Small helper functions can be manually lifted and proven correct relatively easily, and their correctness proofs can be composed together with composing the functions themselves using the theorem. This way, lifted analyses can be compositionally implemented following the same structure of their single-product counterparts, composing correctness proofs together with function composition. We demonstrate this methodology on lifting a Change Impact Assessment (CIA) algorithm in Sec.~\ref{sec:cia}.

\vspace{-0.1in}
\section{Changed Impact Assessment}
\label{sec:cia}
\vspace{-0.1in}
\newcommand{\CIA}{GSN\_IA}

In this section, we formalize the GSN-IA~\cite{Kokaly:2017} impact assessment algorithm, systematically design a lifted version of it, and prove its correctness based on the methodology in Sec.~\ref{sec:methodology}. 

\vspace{-0.1in}
\subsection{Single-Product Algorithm}
\vspace{-0.1in}

\begin{lstlisting}[style=lean, basicstyle=\small, firstline=8,lastline=19,caption=Type definitions of the formalized GSN\_IA algorithm.,label=lst:defs]
inductive Annotation : Type 
| Reuse | Recheck | Revise

constants SysEl GSNEl : Type
def Sys         : Type := set SysEl
def GSN         : Type := set GSNEl
def TraceRel : Type := set (SysEl × GSNEl)

variable sliceSys    (s : Sys)   (es : set SysEl) : Sys
variable sliceGSN_V (ac : GSN) (es : set GSNEl) : GSN
variable sliceGSN_R (ac : GSN) (es : set GSNEl) : GSN
structure Delta := (add : set SysEl) (delete : set SysEl) (modify : set SysEl)
\end{lstlisting}

\newcommand{\Annotation}{\code{Annotation}}
\newcommand{\Reuse}{\code{Reuse}}
\newcommand{\Recheck}{\code{Recheck}}
\newcommand{\Revise}{\code{Revise}}
\newcommand{\SysEl}{\code{SysEl}}
\newcommand{\GSNEl}{\code{GSNEl}}
\newcommand{\Sys}{\code{Sys}}
\newcommand{\GSN}{\code{GSN}}
\newcommand{\TraceRel}{\code{TraceRel}}
\newcommand{\sliceSys}{\code{sliceSys}}
\newcommand{\sliceGSNV}{\code{sliceGSN\_V}}
\newcommand{\sliceGSNR}{\code{sliceGSN\_R}}
\newcommand{\Dlta}{\code{Delta}}
\newcommand{\restrict}{\code{restrict}}
\newcommand{\trace}{\code{trace}}
\newcommand{\createAnnotation}{\code{createAnnotation}}

The data types and external dependencies of the \CIA~algorithm are defined in Listing~\ref{lst:defs}. \Annotation~is the data type of annotations assigned to GSN model elements, with the values \Reuse, \Recheck, and \Revise~(lines 1-2). \SysEl~and \GSNEl~are opaque types of system model elements and GSN model elements respectively, where a system model \Sys~and a GSN model \GSN~ are sets of each of those elements types (lines 4-6). \TraceRel~is a traceability relation between system model elements and GSN model elements, so it is a defined as a set of ordered pairs of \SysEl~and \GSNEl~(line 7). \CIA~is parameterized by three model slicers: \sliceSys~is a system model slicer, while \sliceGSNV~and \sliceGSNR~ are GSN model slicers. 
Each of the slicers takes a model and a set of elements used as the slicing criterion, returning a subset slice of the input model (lines 9-11). \Dlta~is composed of three sets of system elements, representing the elements added, modified and deleted (lines 12).

\begin{lstlisting}[style=lean, basicstyle=\small, firstline=21,lastline=43,caption=Helper functions and the formalized GSN\_IA algorithm.,label=lst:gsnia]
def restrict (t : TraceRel) (d : Delta) : TraceRel :=
    λ x, x.1 ∈ d.add ∪ d.delete ∪ d.modify 

def trace (t : TraceRel) (es : set SysEl) : set GSNEl :=
    image prod.snd {p | p ∈ t ∧ p.1 ∈ es}

def createAnnotation (g : GSN) (recheck : set GSNEl) (revise : set GSNEl)
        : set (GSNEl × Annotation) :=
    let ch := image (λ e, (e, Annotation.Recheck)) recheck,
          rv := image (λ e, (e, Annotation.Revise)) revise,
          ru := image (λ e, (e, Annotation.Reuse)) (g - (recheck ∪ revise))
    in  ch ∪ rv ∪ ru

def GSN_IA (S  : Sys) (S' : Sys) (A  : GSN) (R  : TraceRel) (D  : Delta)
        : set (GSNEl × Annotation) :=
    let R'              := restrict R D,
         C1dm           := sliceSys S ((delete D) ∪ (modify D)),
         C1am           := sliceSys S' ((add D) ∪ (modify D)),
         C2Recheck   := (trace R C1dm) ∪ (trace R' C1am),
         C2Revise     := trace R (delete D),
         C3Recheck1  := sliceGSN_V A C2Revise,
         C3Recheck2  := sliceGSN_R A (C2Recheck ∪ C3Recheck1)
    in  createAnnotation A C3Recheck2 C2Revise
\end{lstlisting}

Listing~\ref{lst:gsnia} has the definitions of the \CIA~algorithm, together with three helper functions. \restrict~ is a function taking a traceability relation ~\code{t}~and a delta ~\code{es}~as inputs, and returns a restricted subset of \code{t} only covering elements in \code{es} (lines 1-2). \trace~takes a traceability relation \code{t}~and a set of system elements~\code{es}~as inputs, and returns the set of GSN elements mapped from \code{es}~by \code{t} (lines 4-5). \createAnnotation~assigns an \Annotation~value to each element in a GSN model, given sets of elements to be rechecked and revised (lines 7-12).


The change impact assessment algorithm \CIA~takes two system models \name{S}~and \name{S'} and the delta \name{D}~between them. It also takes a GSN model \name{A}~and a traceability relation \name{R}~between system model elements and GSN model elements. It returns a set of ordered pairs of GSN model elements and annotations. The algorithm starts by restricting the traceability relation based on \name{D}, slices the original system model \name{S}~using the elements deleted and modified as a slicing criterion, and slices the modified system model \name{S'} using the added and modified elements as the slicing criterion (lines 16-18). Using those two slices, the corresponding GSN model elements are traced using the traceability relation (line 19). The GSN elements traced from elements deleted from the original system model are to be revised (line 20). The slice of the GSN model based on the traced elements are to be rechecked (lines 21-22), and both revise and recheck sets are used to annotate the GSN model elements (line 23). 

\vspace{-0.1in}
\subsection{Lifted Algorithm}
\label{sec:lifting}
\vspace{-0.1in}

\begin{lstlisting}[style=lean,basicstyle=\small,firstline=102,lastline=116,caption=Lifted Change Impact Assessment algorithm.,label=lst:liftedCIA]
def Sys' : Type := set' SysEl
def GSN' : Type := set' GSNEl
def TraceRel' : Type := set' (SysEl × GSNEl)
structure Delta' := (add : set' SysEl) (delete : set' SysEl) (modify : set' SysEl)

def GSN_IA' (S S' : Sys') (A  : GSN') (R  : TraceRel') (D  : Delta')
        : set' (GSNEl × Annotation) :=
    let R'              := restrict' R D,
         C1dm           := sliceSys' S (D.delete ∪ D.modify),
         C1am           := sliceSys' S' (D.add ∪ D.modify),
         C2Recheck   := (trace' R C1dm) ∪ (trace' R' C1am),
         C2Revise     := trace' R D.delete,
         C3Recheck1  := sliceGSN_V' A C2Revise,
         C3Recheck2  := sliceGSN_R' A (C2Recheck ∪ C3Recheck1)
    in  createAnnotation' A C3Recheck2 C2Revise
\end{lstlisting}

Listing~\ref{lst:liftedCIA} is the variability-aware version of the algorithm in Listing~\ref{lst:gsnia}. Both algorithms are compositions of function/operator calls, so each of those functions/operators is replaced with its lifted counterpart. We assume that lifted versions of the three slicers are provided, and that they meet the correctness criteria of Fig.~\ref{fig:correctness1}. 

All the set types used in GSN\_IA need to be lifted. Definitions in lines 1-4 are lifted sets of system model elements, GSN model elements, and traceability mappings. A lifted delta (line 4) is composed of three lifted sets (additions, deletions and modifications).

The proof of the correctness theorem used auxiliary correctness lemmas for each of the helper algorithms. Each of the proofs expands definitions and repeatedly applies the correctness of lifted function composition (Fig.~\ref{fig:correctness2}). 

\vspace{0.1in}
\noindent
{\bf Lifted Helper Algorithms.}
\label{sec:lifted_helpers}
Since the lifted CIA algorithm operates on lifted data structures, all helper algorithms need to be modified to correctly operate on lifted data structures as well. In particular, we outline lifted versions of \restrict~and \trace~(Listing~\ref{lst:liftedHelpers}).

The original implementation of \restrict~takes a traceability map and a delta as inputs, and returns the minimal subset of the traceability map that covers all the elements in the delta. We now have presence conditions associated to system model elements, assurance case elements, and also the traceability links in between. The lifted version of \restrict~(referred to as \lift{\restrict}) needs to correctly process all those presence conditions.

The lifted algorithm starts by calculating the set of relevant elements in the system model, which is the union of added, deleted and modified elements in the delta (line 2). The algorithm returns a lifted traceability mapping as a function taking \code{((s,g),pc)}, where \code{(s,g)} is a system model element-GSN model element pair, and \code{pc} is a presence condition. This function evaluates to the conjunction of applying the input traceability map \code{t} to \code{((s,g),pc)}, and applying \code{relevant} to \code{(s,g)}. Recall that variability-aware sets (as well as Lean sets) are functions mapping values of a given type to propositions.


Similarly, \lift{\trace} is the lifted version of \trace. The returned lifted set is a function mapping a GSN model element \code{g}~to the set of configurations from which there exists a system model element \code{s}~in the input lifted set of system elements, where \code{(s,g)} belongs to the input traceability map.

The lifted version of \createAnnotation~(named \lift{\createAnnotation}) is of exactly the same structure as the original because it strictly uses set operations (union, set difference and image), which have been all lifted as a part of the underlying variability-aware set implementation (Listing~\ref{lst:var}).

\begin{lstlisting}[style=lean,basicstyle=\small,firstline=46,lastline=51,caption=Lifted implementation of \code{restrict} and \code{trace}.,label=lst:liftedHelpers]
def restrict' (t : TraceRel') (d : Delta') : TraceRel' :=
    let relevant := d.add ∪ d.delete ∪ d.modify
    in  λ x, t x ∧ relevant x.1 

def trace' (t : TraceRel') (es : set' SysEl) : set' GSNEl :=
    λ (g:GSNEl), ∃ (s : SysEl) , es s ∧ t ⟨s , g⟩
\end{lstlisting}

%
%
%
%
%

\begin{lstlisting}[style=lean,basicstyle=\small,firstline=147,lastline=149,caption=Correctness theorem of \lift{GSN\_IA}.,label=lst:correctness]
theorem GSN_IA'_correct : 
   ∀ (S S' : Sys') (A  : GSN') (R  : TraceRel') (D : Delta') (pc : PC),
   (GSN_IA' S S' A R D) | pc = GSN_IA  (S | pc) (S' | pc) (A | pc) (R | pc) (D | pc) 
\end{lstlisting}

The correctness theorem of \lift{GSN\_IA} with respect to GSN\_IA is in Listing~\ref{lst:correctness}. It is a direct instantiation of the general correctness criteria in Fig.~\ref{fig:correctness1}, applied to inputs of the GSN\_IA algorithm.

\newcommand{\alarm}{Alarm}
\newcommand{\UI}{UserInterface}
\newcommand{\LDWS}{LaneDepartureWarningSystem}
\newcommand{\audio}{Audio}
\newcommand{\visual}{Visual}

\newcommand{\CtRecheck}{C2recheck}
\newcommand{\CtRevise}{C2revise}
\newcommand{\CrRechecko}{C3recheck1}
\newcommand{\CtRecheckt}{C3recheck2}

\newcommand{\Gfive}{G5}
\newcommand{\Snthree}{Sn3}
\newcommand{\Seight}{S8}
\newcommand{\Geighteen}{G18}
\newcommand{\Gnineteen}{G19}
\newcommand{\Gtwenty}{G20}
\newcommand{\Snfour}{Sn4}
\newcommand{\Sneleven}{Sn11}
\newcommand{\Sntwelve}{Sn12}
\newcommand{\Sneighteen}{Sn18}

\newcommand{\Fldws}{LDWS}
\newcommand{\Faudio}{Audio}
\newcommand{\Fvisual}{Visual}
\newcommand{\Faudiovis}{\Fvisual~$\vee$ \Faudio}

\vspace{-0.2in}
\subsection{Examples}
\vspace{-0.15in}

\begin{figure*}[t]
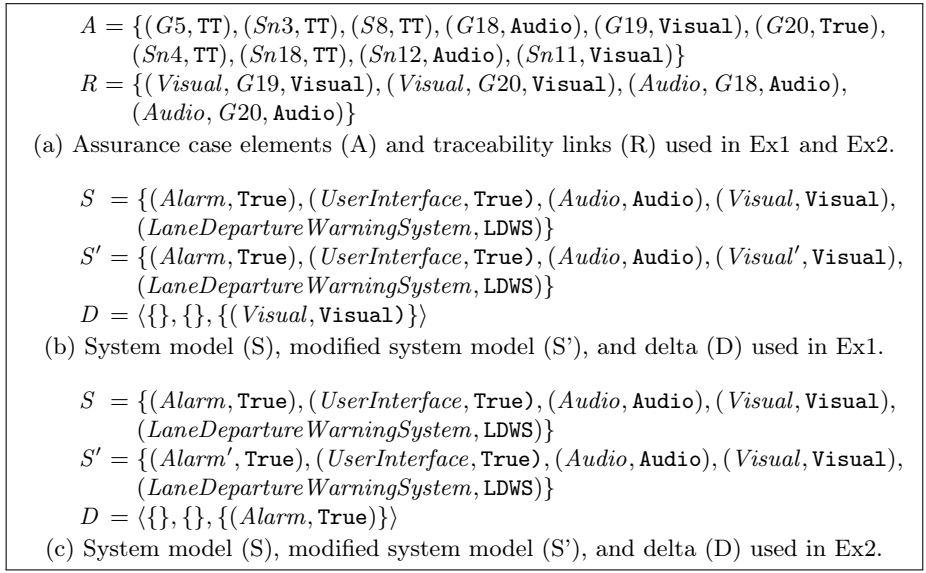

\begin{framed}
\begin{subfigure}[c]{\textwidth}
	\vspace{-0.2in}
	\small
	\centering
	\[
	\begin{array}{lcl}
	A & = & \{ (\Gfive, \PC{TT}), (\Snthree, \PC{TT}), (\Seight, \PC{TT}), (\Geighteen, \PC{\Faudio}), (\Gnineteen, \PC{\Fvisual}), 
	(\Gtwenty, \PC{\TT}), \\
	  &   &   (\Snfour, \PC{TT}), (\Sneighteen, \PC{TT}), (\Sntwelve, \PC{\Faudio}), (\Sneleven, \PC{\Fvisual}) \} \\
	
	R & = & \{ (\visual, \Gnineteen, \PC{\Fvisual}), (\visual, \Gtwenty, \PC{\Fvisual}), (\audio, \Geighteen, \PC{\Faudio}), \\
	  &   &   (\audio, \Gtwenty, \PC{\Faudio}) \}
	\end{array}
	\]
	\vspace{-0.2in}
	\caption{Assurance case elements (A) and traceability links (R) used in Ex1 and Ex2.}
	\label{fig:common}
\end{subfigure}

\vfill

\begin{subfigure}[c]{\textwidth}
	\small
	\centering
	\[
	\begin{array}{lcl}
	S & = & \{ (\alarm, \PC{\TT}), (\UI, \PC{\TT)}, (\audio, \PC{\Faudio}), (\visual, \PC{\Fvisual}), \\
	  &   &     (\LDWS, \PC{\Fldws}) \} \\

	S' & = & \{ (\alarm, \PC{\TT}), (\UI, \PC{\TT)}, (\audio, \PC{\Faudio}), (\visual', \PC{\Fvisual}), \\ 
	   &   &    (\LDWS, \PC{\Fldws})\} \\

	D & = & \langle \{\}, \{\}, \{(\visual, \PC{\Fvisual)}\} \rangle
	\end{array}
	\]
	\vspace{-0.2in}
	\caption{System model (S), modified system model (S'), and delta (D) used in Ex1.}
	\label{fig:example1}
\end{subfigure}

\vfill

\begin{subfigure}[c]{\textwidth}
	\small
	\centering
	\[
	\begin{array}{lcl}
	S & = & \{ (\alarm, \PC{\TT}), (\UI, \PC{\TT)}, (\audio, \PC{\Faudio}), (\visual, \PC{\Fvisual}), \\
	  &   &    (\LDWS, \PC{\Fldws})\} \\
	
	S' & = & \{ (\alarm', \PC{\TT}), (\UI, \PC{\TT)}, (\audio, \PC{\Faudio}), (\visual, \PC{\Fvisual}), \\ 
	   &   &   (\LDWS, \PC{\Fldws})\} \\
	
	D & = & \langle \{\}, \{\}, \{(\alarm, \PC{\TT})\} \rangle
	\end{array}
	\]
	\vspace{-0.2in}
	\caption{System model (S), modified system model (S'), and delta (D) used in Ex2.} 
	\vspace{-0.1in}
	\label{fig:example2}
\end{subfigure}
\end{framed} 
\vspace{-0.2in}
\caption{Inputs to the \lift{GSN-IA} algorithm used in Ex1 and Ex2.}
\label{fig:inputs}
\vspace{-0.2in}
\end{figure*}

In this section, we apply our lifted CIA algorithm to two examples of modifications to the fragment of the LMS product line presented in Sec.~\ref{sec:intro} (Fig.~\ref{fig:ex}). 

\vspace{0.1in}
\noindent
{\bf Ex1:  Feature-Specific Modification.}
Suppose  that the \term{\visual} class is modified. This class is local to the \PC{\Fvisual} feature. If we only analyze the fragment in Fig.~\ref{fig:ex}, the inputs to \lift{GSN-IA} are shown in \fig{common} and \fig{example1}.


Tracing through the algorithm, the first step is using \lift{\restrict} to calculate \Rn = \{(\visual, \Gnineteen, \PC{\Fvisual}), (\visual, \Gtwenty, \PC{\Fvisual})\} (line 8). Because \Cza~and \Czd~are both empty, and assuming a backward slicer (returning the transitive closure of the elements that might affect the slicing criteria), \Codm~and \Coam~both become \{(\alarm, \PC{\TT}), (\visual, \PC{\Fvisual}), (\LDWS, \PC{\Fldws})\} (lines 9-10). Now tracing from \Codm~and \Coam, \CtRecheck~becomes \{(\Gnineteen, \PC{\Fvisual}), (\Gtwenty, \PC{\Fvisual})\} (line 11).
Since \Czd~is empty, \CtRevise~and \CrRechecko~ are both empty as well (lines 12-13).  Using a backward GSN slicer, 
\CtRecheckt~becomes \{(\Gnineteen, \PC{\Fvisual}), (\Gtwenty, \PC{\Fvisual}), (\Sneleven, \PC{\Fvisual}), (\Snfour, \PC{\Fvisual}), (\Sneighteen, \PC{\Fvisual})\} (line 14). The algorithm returns an empty set of GSN elements to be revised, and the set \CtRecheckt~to be rechecked. Note that \Gtwenty, \Snfour, and \Sneighteen~are all base model elements (having \PC{\TT} as a presence condition), so the algorithm output states that we need to recheck those elements only in products where the feature \PC{\Fvisual}~is present.
 
 \vspace{0.1in}
 \noindent
 {\bf Ex2:  Base System Modification.}
\label{sec:example2}
Suppose that the \term{\alarm} class is modified. This is a base system class, i.e., it is present in all products. The inputs to \lift{GSN-IA} (restricted to the fragment in Fig.~\ref{fig:ex}) are shown in \fig{common} and \fig{example2}.

Since the \alarm~class does not have any direct traceability links, $\Rn$ is empty (line 8). Using a backward slicer (like in Ex1), \Codm~and \Coam~both become \{(\alarm, \PC{\TT}), (\visual, \PC{\Fvisual}), (\audio, \PC{\Faudio}), (\LDWS, \PC{\Fldws})\} (lines 9-10). 
From \Codm~and \Coam~using the traceability links, \CtRecheck~becomes \{(\Geighteen, \PC{\Faudio}), (\Gnineteen, \PC{\Fvisual}), (\Gtwenty, \PC{\Fvisual})\} (line 11).
Again, since \Czd~is empty, \CtRevise~and \CrRechecko~ are both empty as well (lines 12-13). With a backward GSN slicer, \CtRecheckt~becomes \{(\Geighteen, \PC{\Faudio}), (\Gnineteen, \PC{\Fvisual}), (\Gtwenty, \PC{\Faudiovis}), (\Sneleven, \PC{\Fvisual}), (\Sntwelve, \PC{\Faudio}), (\Snfour, \PC{\Faudiovis}), (\Sneighteen, \PC{\Faudiovis})\} (line 14). 
The algorithm returns an empty set of GSN elements to be revised, and the set \CtRecheckt~to be rechecked. 
Note that in this example, \Gtwenty, \Snfour, and \Sneighteen~are annotated with \name{recheck} with presence condition \PC{\Faudiovis}, which means that they need to be rechecked only if either \PC{\Faudio} or \PC{\Fvisual} are present. 

\vspace{-0.1in}
\section{Towards Implementation}
\label{sec:impl}
\vspace{-0.1in}

\begin{figure}[t]
	\includegraphics[width=\textwidth]{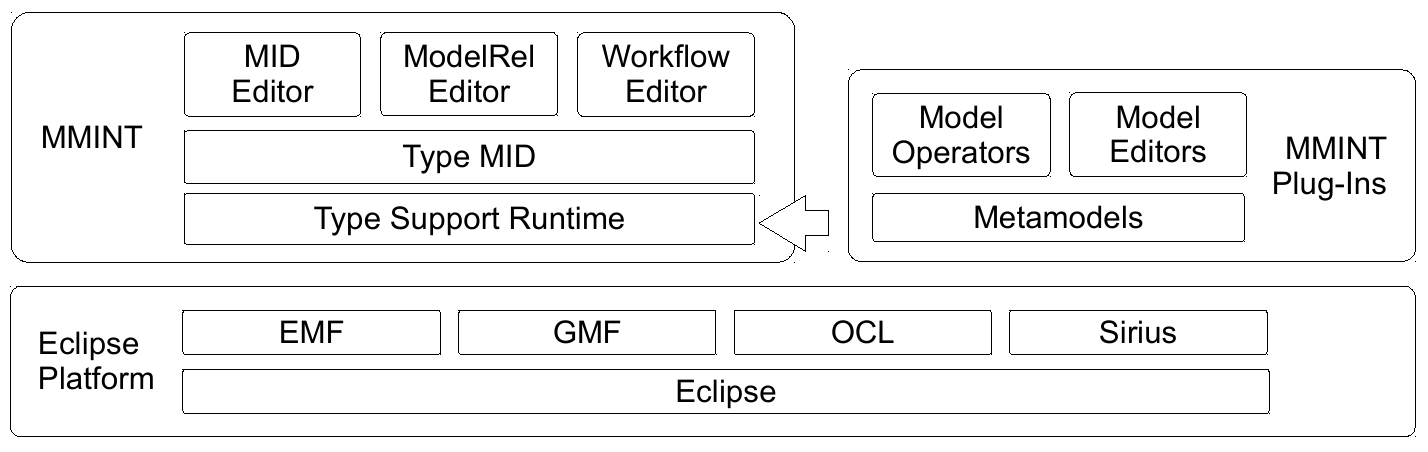}	
	\vspace{-0.2in}
	\caption{Architecture of the MMINT model management framework~\cite{Fung:2018}.}
	\label{fig:mmint}	
	\vspace{-0.2in}
\end{figure}

The GSN-IA algorithm is implemented, together with slicers and model operators, as an extension of the MMINT~\cite{DiSandro:2015} model management framework (Fig.~\ref{fig:mmint}), called MMINT-A~\cite{Fung:2018}. In order to extend MMINT-A to support annotative product line models, and subsequently the lifted change impact assessment algorithm, the following modifications are required:
(1) Model elements need to be extended with presence conditions, with $\TT$ as a default value.   This way single product models (where all elements have the default $\TT$ presence condition) are directly supported as well.
(2) Operators on models need to be modified to take presence conditions into consideration, and compute the presence conditions of their outputs. Those modifications are mostly systematic along the lines of those of \lift{\restrict} and \lift{\trace} (Listing~\ref{lst:liftedHelpers}). 
(3) Higher-level algorithms (e.g., GSN-IA) need to be modified accordingly to use the lifted versions of the operators.
(4) The user interface of MMINT-A needs to support annotating different model elements with presence conditions.
(5) Optionally, MMINT-A can check the well-formedness of presence condition annotations. For example, the presence condition of an association between two UML classes has to be subsumed by the presence conditions of its two end points.

\vspace{-0.1in}
\section{Related Work}
\label{sec:related}
\vspace{-0.1in}
\noindent
{\bf Model-based approaches to safety case management.}  Many methods
for modeling safety cases have been proposed, including
goal models and requirements models~\cite{Ghanavati:2011,Brunel:2012} and GSN~\cite{Kelly:2004}.  The latter is
arguably the most widely used model-based
approach to improving the structure of safety arguments.  Building on GSN,
Habli et. al.~\cite{Habli:2010} examine how model-driven development can
provide a basis for the systematic generation of functional safety
requirements and demonstrates how an automotive safety case can be developed.  Gallina~\cite{Gallina:2014}
proposes a model-driven safety certification method to derive
arguments as goal structures given in GSN from process models. The
process is illustrated by generating arguments in the context of ISO
26262. We consider this category of work complimentary to ours; we do
not focus on safety case construction but instead
assume presence of a safety case
and focus on assessing the impact of system changes.


\vskip 0.1in
\noindent{\bf Lifting to Software Product Lines.} Different kinds of software analyses have been re-implemented to support product lines~\cite{Thum:2014}. 
For example, the TypeChef project implements variability aware parsers and type checkers for Java and C~\cite{Kastner:2012}. The SuperC project~\cite{Gazzillo:2012} is another C language variability-aware parser. A graph transformation engine was lifted to product lines of graphs~\cite{Salay:2014}. Datalog-based analyses (e.g., pointer analysis) have been lifted by modifying the Datalog engine being used~\cite{Shahin:2019}.
SPL\textsuperscript{Lift}~\cite{Bodden:2013} lifts data flow analyses to annotative product lines. Model checkers based on Featured Transition Systems~\cite{Classen:2013} check temporal properties of transition systems where transitions can be labeled by presence conditions. 
Syntactic transformation techniques have been suggested for lifting abstract interpretation analyses~\cite{Midtgaard:2015} and functional analyses~\cite{Shahin:2020} to SPLs.

In this paper, our methodology tailors the lifting approach from related work to safety cases of product lines, and we demonstrate it on change impact assessment. 
We tackle a new class of product line artifacts, particularly safety cases. To the best of our knowledge, this is the first attempt to lift a safety case analysis to product lines.
 
\vskip 0.1in
\noindent{\bf Formalized Systems and Interactive Theorem Proving.} Correctness and behavioral properties of several software systems have been formalized and verified using interactive theorem provers. The CompCert compiler~\cite{Leroy:2009} is an example of a C-language compiler fully certified using the Coq theorem prover. The seL4 microkernel~\cite{Klein:2009} was verified using the Isabelle\textbackslash HOL theorem prover. Isabelle was also used to formalize the Structured Assurance Case Metamodel (SACM) notation for certified definition of assurance cases~\cite{Nemouchi:2019}. 
\vspace{-0.1in}
\section{Conclusion and Future Work}
\label{sec:conclusion}
\vspace{-0.1in}
 In this paper, we presented a methodology for lifting safety case analysis algorithms to software product lines. We also outlined a certification infrastructure (data structures and correctness criteria) for our lifting approach using the Lean interactive theorem prover. We demonstrated both the approach and correctness certification on formalizing and lifting a Change Impact Assessment (CIA) algorithm~\cite{Kokaly:2017}. We discussed the implementation of the lifted CIA algorithm as part of the safety model management system MMINT-A~\cite{Fung:2018}. 
 A lifted CIA algorithm allows for reusing impact assessment conclusions across a potentially exponential  (in number of features) different product variants, as opposed to using a product-level CIA algorithm in individual product instances, which is intractable in most cases.
 
 For future work, we are working together with an industrial partner on applying our lifted algorithm to their assurance case models. We also plan to lift other safety case algorithms (including slicers), and add their implementations to MMINT-A. Visualization of the analysis results and improved user interaction is another area of future improvements.
 \vspace{-0.1in}

\bibliographystyle{splncs04}
\bibliography{spl,modeling,pl,safety}
\end{document}